\documentclass[journal=jacsat,reprint,manuscript=article]{achemso}
\usepackage{mathrsfs}
\usepackage{comment}
\usepackage[caption=false]{subfig}
\usepackage{verbatim}
\usepackage[utf8]{inputenc}
\usepackage{textcomp}
\usepackage{lineno}
\usepackage{amsmath,textcomp,gensymb}
\usepackage{amsfonts}
\usepackage{amssymb}
\usepackage{amsthm}
\usepackage{graphicx}
\usepackage{natbib}
\usepackage{color}
\usepackage{pdfpages}
\usepackage{hyperref}
\usepackage{bm}

\usepackage{soul}
\usepackage[normalem]{ulem}
\usepackage[version=3]{mhchem}

\author{Sohini Guin$^{\dagger}$}
\affiliation{Centre for Nanoscience and Engineering, Indian Institute of Science, Bengaluru, Karnataka 560012, India} 
\author{Ramesh Kumar Guin$^{\dagger}$}
\affiliation{Centre for Nanoscience and Engineering, Indian Institute of Science, Bengaluru, Karnataka 560012, India} 
\author{Pankaj Bhardwaj}
\affiliation{Centre for Nanoscience and Engineering, Indian Institute of Science, Bengaluru, Karnataka 560012, India} 
\author{Laxmipriya Nanda}
\affiliation{Centre for Nanoscience and Engineering, Indian Institute of Science, Bengaluru, Karnataka 560012, India} 
\author{Rajib~Sarkar}
\affiliation{Centre for Nanoscience and Engineering, Indian Institute of Science, Bengaluru, Karnataka 560012, India}
\author{Dhavala Suri}
\email{dsuri@iisc.ac.in}
\affiliation{Centre for Nanoscience and Engineering, Indian Institute of Science, Bengaluru, Karnataka 560012, India}

\title{From Growth to Devices: Oxygen-Controlled Superconductivity in BSCCO Thin Films}

\begin{document}

\begin{abstract}
Realizing device-compatible high-$T_c$ superconductors requires simultaneous control of oxygen stoichiometry and tolerance to nanofabrication. Here, we investigate sputtered Bi$_2$Sr$_2$Ca$_{n-1}$Cu$_n$O$_{2n+4+\delta}$ (BSCCO) thin films and establish a direct link between annealing conditions, oxygen incorporation, and superconducting transport. By independently varying the annealing temperature and duration, we identify an optimal annealing condition of $860\,^{\circ}$C for 3\,h, which yields the highest superconducting transition temperature and the narrowest transition among the investigated samples. X-ray diffraction, electron microscopy, and X-ray photoemission spectroscopy reveal a non-monotonic evolution of lattice oxygen, which governs hole doping in the CuO$_2$ planes and produces the characteristic dome-like dependence of $T_c$ on doping concentration.  We further translate these optimized films into micro-devices to assess fabrication-induced effects, via micro-bridges as well as a planar Josephson junction. We demonstrate devices fabricated via both lift-off and  focused ion beam (FIB) patterning methods -- proving the fabrication compatibility of these films. Importantly, we demonstrate their air-stability for extended periods of time (2 months). The combined insights offer practical guidelines for integrating cuprate thin films into superconducting electronics and quantum device architectures.

\end{abstract}

\maketitle

High-quality thin films of high-temperature superconductors are essential for Josephson junctions, bolometers, and quantum sensors~\cite{Likharev1979,Clarke2004,Charaev2023,Gupta2018,Braginski1992,Couedo2020}. Among cuprate superconductors, Bi$_2$Sr$_2$CaCu$_2$O$_{8+\delta}$ (BSCCO, also referred to as Bi-2212) is particularly attractive for its high transition temperature and suitability for probing low-dimensional superconductivity~\cite{Maeda1988,Damascelli2003,Keimer2015}. Recent studies on twisted and atomically thin BSCCO have demonstrated the potential of exfoliated cuprate flakes for studying unconventional superconductivity and Josephson physics~\cite{Jindal2017,Sterpetti2017,Liao2018,Yu2019,Wang2021,Lei2022,Zhao2023,Martini2023,Lee2023,Confalone2025,Zhao2019,Ghosh2024}. While early work showed that annealing could establish Josephson coupling between stacked BSCCO crystals~\cite{Li1999}, more recent studies revealed the sensitivity of these interfaces to disorder, oxygen stoichiometry, and fabrication conditions. To preserve interface quality and Josephson coupling, cryogenic stacking and encapsulation techniques have therefore been developed~\cite{Zhao2023,Martini2023,Lee2023,Confalone2025}, and low-temperature doping studies further highlight the sensitivity of atomically thin BSCCO to processing~\cite{Yu2019,Zhao2019}. Although flake-based approaches provide an important route to the intrinsic properties of high-quality crystals, their reliance on exfoliation, deterministic stacking, and specialized processing limits reproducible and scalable device fabrication. Continuous BSCCO thin films thus offer an attractive route toward lithographically defined, scalable high-temperature superconducting devices.

Yet synthesizing BSCCO thin films remains challenging due to multicomponent chemistry, a narrow thermodynamic stability window, and strong sensitivity to oxygen stoichiometry~\cite{Balestrino2005,Eisaki2004}. Among physical vapor deposition techniques, pulsed laser deposition (PLD) yields high crystallinity, molecular beam epitaxy (MBE) enables atomic-layer control, and sputtering offers scalability and device fabrication compatibility~\cite{Eom1992,Bozovic2002,Krockenberger2008,keppert2023phase,Mattox2010,Rossnagel2003,Nane2013,Nane2016}. However, sputtered films often exhibit non-equilibrium stoichiometry, limited adatom mobility, and secondary phase formation~\cite{Balestrino2005,Koster2012}. Post-deposition annealing is therefore critical for driving the system toward thermodynamic equilibrium. Annealing temperature and duration jointly govern cation and oxygen diffusion kinetics, phase stability, and crystallographic ordering~\cite{Deville1997,Venables1984,Zhang1993}. At insufficient annealing temperatures or durations, kinetic barriers suppress long-range diffusion, yielding metastable phases and compositional disorder. At optimal conditions, enhanced atomic mobility promotes formation of the thermodynamically stable BSCCO phase, improving crystallinity and texture. Excessively high temperatures, however, cause bismuth volatility, film--substrate interdiffusion, and surface roughening driven by minimization of surface and interfacial free energies~\cite{Balestrino2005,Koster2012}. Together, annealing temperature and duration define a practical growth window bounded by thermodynamic stability and kinetic accessibility. Despite the extensive literature on bulk and single-crystal BSCCO \cite{Wen1993,Tanaka2000,Tanaka2001,Tanaka2002,Funabiki2003,Ishida2004,Imao2005,Adachi2007,Echizen2009,Kumar2004,Kumar2018}, systematic studies correlating thin-film growth conditions with crystallinity, oxygen stoichiometry, superconducting properties, and device compatibility remain scarce, and device-level demonstrations are sparser still.

Developing a reproducible BSCCO thin-film platform is therefore essential for scalable high-temperature superconducting devices and circuits. In contrast to flake-based approaches, this work establishes explicit processing--oxygen--superconductivity correlations in sputtered, device-compatible BSCCO thin films, and critically assesses the fabrication-induced degradation pathways relevant for device integration. By independently tuning annealing temperature and duration, we identify a narrow kinetic window that simultaneously optimizes phase formation, crystallographic texture, and oxygen stoichiometry. Structural, chemical, and transport measurements reveal that superconducting properties exhibit a non-monotonic dependence on lattice oxygen content, which directly governs hole doping in the CuO$_2$ planes. Importantly, we extend this optimization to the device scale using both lift-off and focused ion beam (FIB) patterning  techniques. While FIB enables high-resolution definition of micron- and sub-micron-scale geometries, the energetic ion bombardment may introduce localized damage, including collision-induced defect generation, amorphization, and preferential sputtering of oxygen, all of which can perturb the delicate oxygen stoichiometry of cuprates. Therefore, an alternate method is to fabricate micro-devices from films grown within the optimized processing window via lift-off. We attempt both techniques in this work to prove the device compatibility of these sputter grown films.

Thin-film growth of BSCCO by sputtering involves several interdependent parameters, including deposition pressure, substrate temperature, target-to-substrate distance, gas composition and flow rate, and post-deposition annealing conditions. These parameters can influence the oxygen stoichiometry of the film, which plays an important role in determining its superconducting properties, particularly the transition temperature $T_c$. In this work, we focus on post-deposition annealing as a controlled route for tuning the oxygen stoichiometry and systematically investigate the effects of annealing temperature and duration while keeping the other deposition parameters fixed. 

\begin{figure*}[t]
\centering
\includegraphics[width=0.7\textwidth]{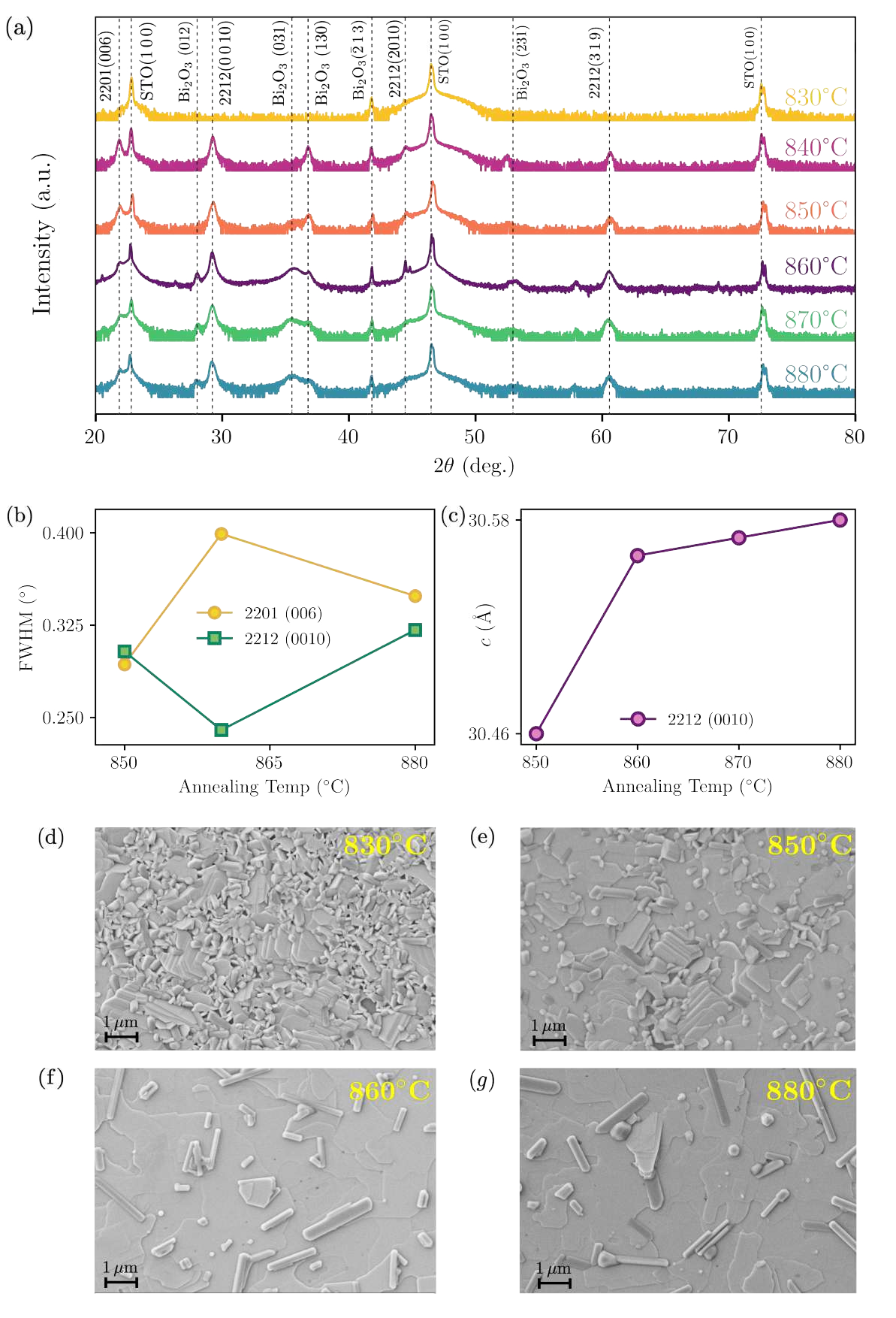}
\caption{Structural and morphological evolution of $\approx 110~\mathrm{nm}$-thick BSCCO films as a function of annealing temperature. 
(a) X-ray diffraction (XRD) patterns for films annealed from 830$^\circ$C to 880$^\circ$C by keeping same annealing durations of 3 hours, showing the evolution from mixed-phase, broadened features at lower temperatures to sharp, c-axis oriented (0010) reflections characteristic of the 2212 phase at intermediate temperatures, followed by degradation at higher temperatures. 
(b) Full width at half maximum (FWHM) of the (i) 2212 (0010) and (ii) 2201 (006) reflections as a function of annealing temperature, highlighting optimal crystalline coherence near 850–860$^\circ$C and competing phase formation. (c) Lattice constant (\textit{c}) of the 2212 phase with the evolution of annealing temperature, showing lattice expansions, confirming the effects of thermal expansions. (d–g) Scanning electron microscopy (SEM) images of films annealed at (d) 830$^\circ$C, (e) 850$^\circ$C, (f) 860$^\circ$C, and (g) 880$^\circ$C, illustrating the transition from dense, disordered grains to well-defined plate-like crystallites and subsequent coarsening and reduced surface coverage at higher temperatures.}
\label{temp}
\end{figure*}

Films were deposited at room temperature, at a base vacuum of $\approx~5~\times$ 10$^{-8}$~mbar, 100~W RF power, 40~sccm Ar flow, and a process pressure of $5\times10^{-3}$~mbar on STO-100 substrates, across a range of thicknesses. The deposited films were subsequently annealed in an O$_\text{2}$ atmosphere at varying temperatures and durations [see SI for details]. XRD and SEM analyses (Fig.~1) reveal a transition from mixed-phase, disordered films at low annealing temperature to highly c-axis-oriented 2212 phase at intermediate temperatures, followed by degradation at higher temperatures \cite{Nane2013}. The (00$10$) reflections confirm phase evolution of BSCCO-2212 phase, with optimal crystallinity near 850–860$^\circ$C as indicated by minimum FWHM of 0.239$^\circ$ 
[Fig. 1~(b)]. The presence of peaks (00\textit{l}) at approximately 21.9$^\circ$  and 29.2$^\circ$ corresponds to the peaks (006) and (0010) of the BSCCO 2201 and 2212 phases, respectively, having lattice parameters ‘\textit{c}’ of 24.276~\AA~and 30.56 \AA, confirming the single and double planes of CuO$_2$ in sputtered BSCCO thin films \cite{Gayathri2021}. In Fig. 1(c) with increasing annealing temperature, the BSCCO-2212 phase exhibits an increase in the lattice constant `c' from  30.46~\AA to 30.58~\AA, confirming the development of tensile strain due to thermal expansion and suggesting the onset of structural defects and disorder. SEM images show concurrent morphological evolution from fine grains to plate-like crystallites and eventual coarsening due to bismuth volatility [Fig. 1(d-g)]. These results identify an optimal temperature window as $\approx$ 860$^{\circ}$ balancing phase formation and structural stability. As seen in the SEM image, the rectangular rod like features maybe attributed to manifestations of Bi rich phases (via XRD spectrum): either 2201 phase of BSCCO or Bi$_2$O$_3$.

\begin{figure}[t]
\includegraphics[width=0.65\columnwidth]{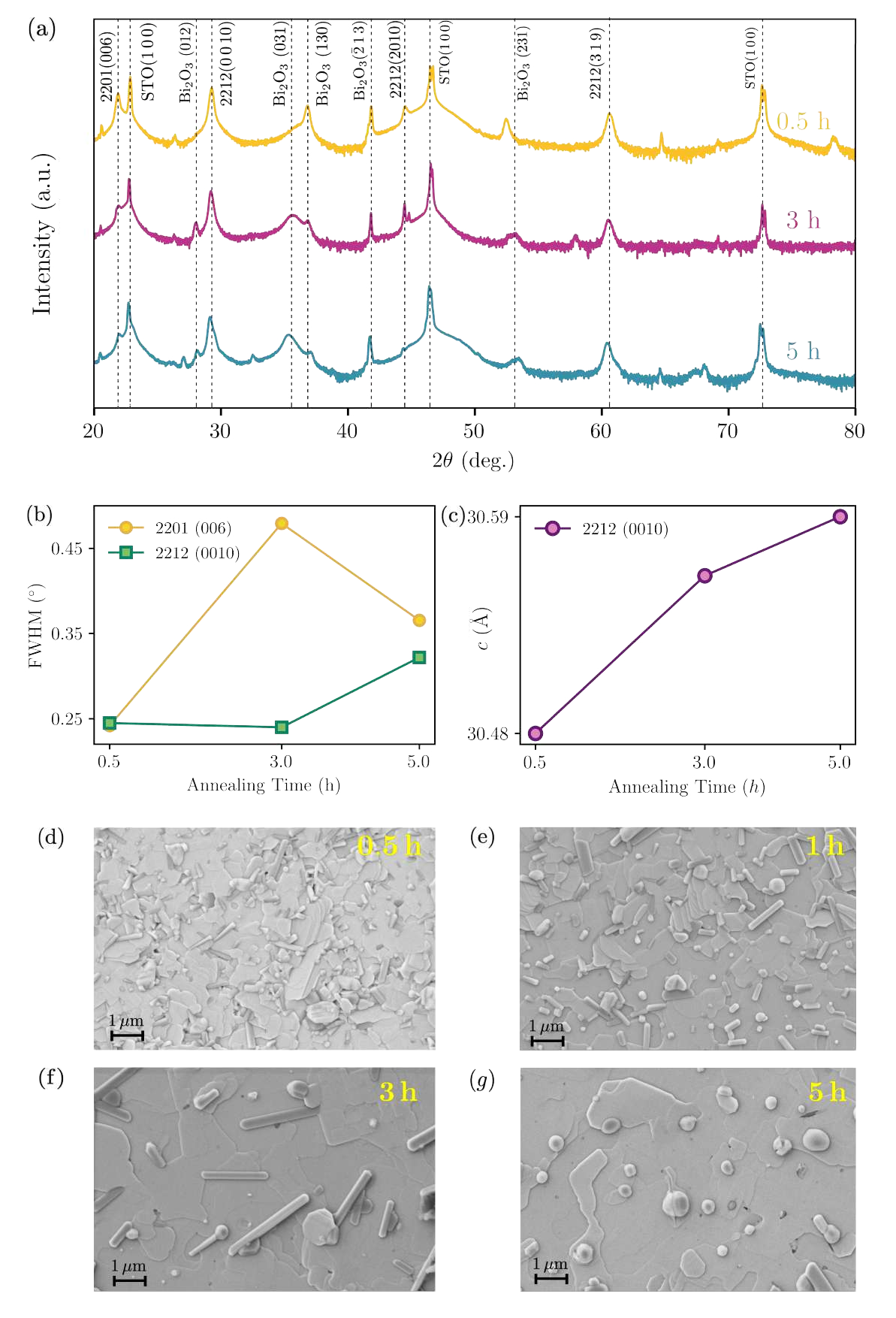}
\caption{Influence of Annealing Duration on the Structural and Morphological Properties of $\approx 110~\mathrm{nm}$-thick BSCCO films
 at a fixed temperature of 860$^\circ$C. (a) X-ray diffraction (XRD) patterns for films annealed for 0.5 h, 3 h, and 5 h, showing the evolution from broadened, weak reflections at short durations to sharp and intense (00$l$) peaks indicative of well-developed 2212 phase and c-axis orientation at intermediate times, followed by degradation at longer durations. (b) Full width at half maximum (FWHM) of the (i) 2212 (0010) and (ii) 2201 (006) reflections as a function of annealing time. (c) Lattice constants (\textit{c}) show the expansions with an increase in annealing durations. (d–g) Scanning electron microscopy (SEM) images for films annealed for (d) 30 min, (e) 1 h, (f) 3 h, and (g) 5 h, illustrating the transition from nucleation-dominated, fine-grained morphology to well-defined plate-like crystallites at intermediate times, and subsequent grain coarsening, reduced surface coverage, and morphological degradation at prolonged annealing.  }
\label{time}
\end{figure}

To further elucidate the role of growth kinetics at the optimized annealing temperature, Fig.~2 examines the evolution of structural and morphological properties as a function of annealing duration at 860$^\circ$C. The X-ray diffraction patterns in Fig.~2(a) show a clear progression from relatively broad and weak reflections at shorter durations (30 min) to sharper and more intense peaks at intermediate times, followed by a degradation at longer annealing times. At 30 min, incomplete crystallization and residual disorder dominate, as evidenced by peak broadening and the presence of multiple reflections, indicating that the system has not reached equilibrium phase formation. Increasing the annealing time to 3 h results in significant sharpening of the (00$l$) peaks associated with the 2212 phase, reflecting enhanced atomic diffusion, improved phase percentage, and the development of c-axis oriented growth. This trend is quantitatively supported by the FWHM analysis in Fig.~2(b), where the (0010) reflection exhibits a minimum at intermediate annealing time, indicating optimal crystalline coherence and reduced microstrain. Concurrently, the (006) peak associated with the 2201 phase shows a non-monotonic variation, suggesting a transient competition between phase formation and intergrowth stabilization.

It is also observed that, with increasing annealing time, the lattice parameters of the 2201 phase contract [see SI], whereas the 2212 phase expands. As the annealing time increases, the 2201 phase experiences greater compressive stress, whereas the 2212 phase experiences tensile strain, as observed in Fig. 2(c) \cite{Mark2023}. At longer annealing durations (5 h), the re-broadening of diffraction peaks and the reduction in peak intensity point to the onset of structural degradation, likely driven by prolonged thermal exposure leading to bismuth volatility, defect accumulation, and possible decomposition or phase segregation. The SEM images in Fig. 2~(d-g) closely mirror these structural trends. At short annealing times (30 min), the morphology is characterized by small, densely packed, and irregular grains, indicative of nucleation-dominated growth. As the annealing duration increases to 1 h and 3 h, well-defined plate-like and elongated grains emerge, reflecting anisotropic growth along the CuO$_2$ planes facilitated by enhanced surface diffusion and layer-by-layer stacking. However, at 5 h, the morphology becomes increasingly coarsened, with isolated large grains and reduced surface coverage, suggesting re-evaporation and mass loss effects. This coarsening is accompanied by the appearance of rounded features, indicative of surface energy minimization and possible partial melting or recrystallization processes. We infer that 3 h annealing duration serves as optimal duration window for the optimal temperature (860$^\circ$C). 

Taken together with the temperature-dependent results in Fig.~1, these observations establish that both annealing temperature and duration must be carefully optimized to balance phase formation, defect minimization, and compositional stability. While insufficient annealing time limits crystallization, excessive duration leads to structural degradation, highlighting a narrow processing window where crystalline quality, phase purity, and microstructural uniformity are maximized, which is expected to be critical for achieving optimal superconducting performance. Our optimal parameters are 860$^{\circ}$C for 3 h. It is important to note that annealing alone does not guarantee microscopic chemical uniformity, phase purity, an atomically homogeneous surface, or an optimal interface for device fabrication.

\begin{figure*}[t]
\centering
\includegraphics[width=\textwidth]{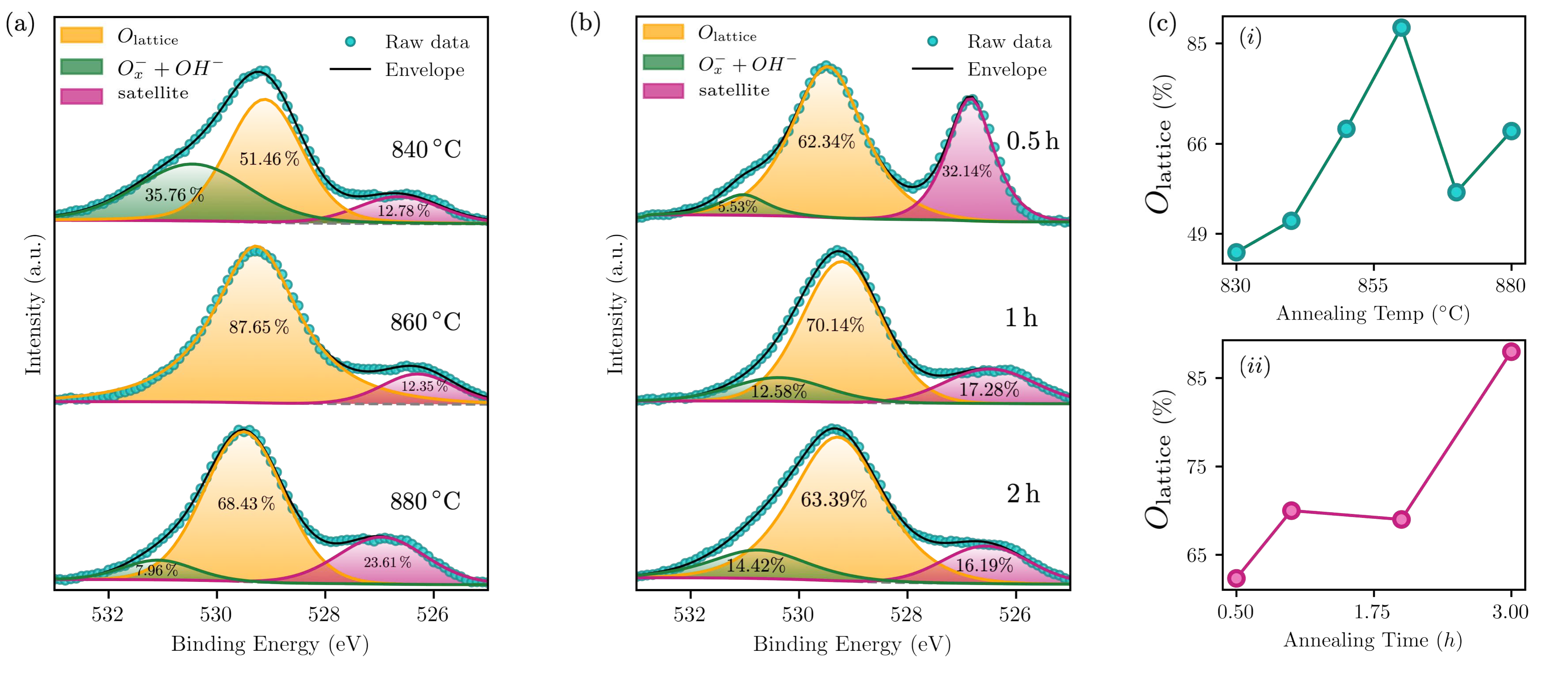}
\caption{X-ray photoemission spectroscopy (XPS) analysis of oxygen states in BSCCO thin films, illustrating the evolution of oxygen stoichiometry with annealing conditions. (a) Deconvoluted O 1$s$ spectra for films annealed at 840$^\circ$C, 860$^\circ$C, and 880$^\circ$C  (all annealed for 3 h), showing contributions from lattice oxygen (O$_{\mathrm{lattice}}$), defect-related or non-stoichiometric oxygen (O$_x^-$) with hydroxyl contributions (OH$^-$) and an additional residual X-ray satellite peak. (b) O 1$s$ spectra for films annealed at 860$^\circ$C for varying durations (0.5 h, 1 h, and 2 h), revealing enhanced lattice oxygen contribution at intermediate annealing times. (c) Extracted fraction of lattice oxygen percentage (O$_{\mathrm{lattice}}$) plotted as a function of (i) annealing temperature and (ii) annealing time, exhibiting a non-monotonic dependence that defines an optimal processing window for stoichiometric incorporation. Note: all subfigures present data from samples with a thickness of $\approx 110~\mathrm{nm}$.}
\label{xps}
\end{figure*}

X-ray photoemission spectroscopy (XPS) was employed to determine the elemental composition and chemical oxidation states of the near surface region of sputtered BSCCO thin films. Specifics regarding the XPS methodology are provided in Section 9 of the SI. XPS analysis revealed the variation in oxygen (Fig.~3), bismuth (SI S10, S11), and copper (SI S10, S11) stoichiometry as a function of annealing conditions. The O 1s spectra comprises contributions from stoichiometric, non-stoichiometric lattice oxygen, as well as hydroxyl components. The binding energy (BE) positions of the non-stoichiometric lattice oxygen, and hydroxyl components are very proximate to each other, and occur at higher BE, which complicates their contributions. However, higher annealing temprature during synthesis and $\text{Ar}^+$ etching reduce the possibility of hydroxyl groups. Consequently, the O 1s spectra are deconvoluted into lattice oxygen (O$_{\text{lattice}}$) and non-stoichiometric components (O$_x^-$) with surface hydroxyl (OH$^-$) impurities having BE centered around $\sim$529.8~eV and $\sim$531.5 eV, respectively with additional peak positioned at $527 \pm 0.5$~eV occurred because of characteristic of a residual X-ray satellite associated with
minor K$\alpha_{3,4}$ emission lines that are not fully suppressed by the
monochromator \cite{Nichols2014}, as shown in Fig.~3~(a) and~(b). The lattice oxygen fraction exhibits a non-monotonic dependence on both annealing temperature and duration, peaking near optimal growth conditions [Fig. 3(c)]. This trend directly correlates with improved crystallinity from XRD, indicating that oxygen incorporation governs structural ordering and defect reduction. At higher temperatures or prolonged annealing, reduced lattice oxygen reflects oxygen loss and compositional instability [see SI for details]. Importantly, the relative increase in lattice oxygen is accompanied by a reduction in defect-related oxygen components, indicating improved ordering within the CuO$_2$ planes and reduced disorder, at 860$^\circ$C for 3h annealing. This has direct implications for superconducting properties, as oxygen stoichiometry governs the hole concentration in the CuO$_2$ layers. 

\begin{figure*}[t]
\centering
\includegraphics[width=\textwidth]{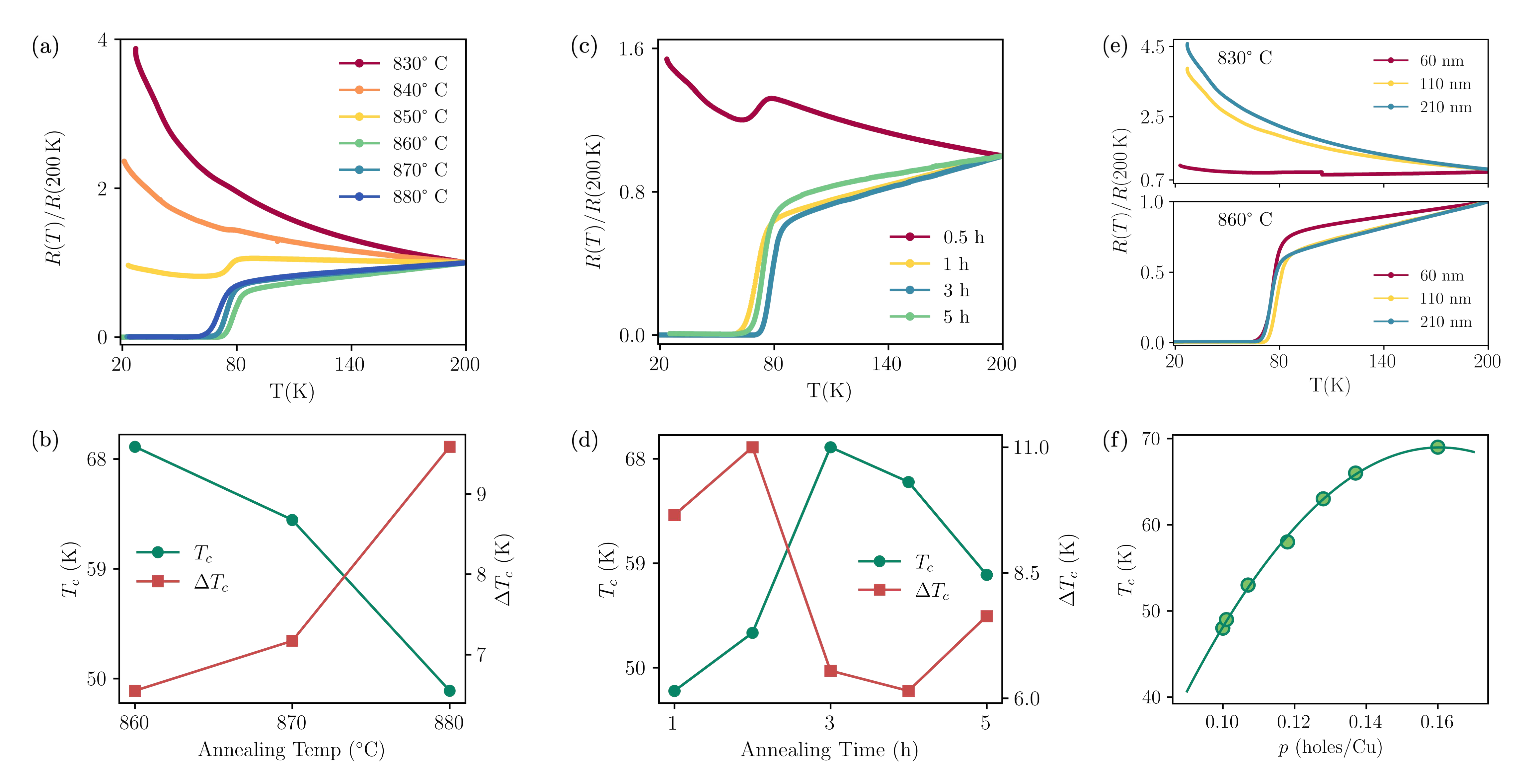}
\caption{Superconducting transport properties and carrier concentration of BSCCO thin films as a function of annealing conditions. (a) Normalized resistance $R(T)/R(200\,\mathrm{K})$ for films annealed at different temperatures, showing a systematic evolution of the superconducting transition, with the sharpest transition and lowest normalized resistance observed near 860$^\circ$C, indicative of optimal structural and stoichiometric conditions. (b) Superconducting transition temperature ($T_c$) and extracted $\Delta T_c$ as a function of annealing temperature. (c) Normalized resistance for films annealed at 860$^\circ$C for varying durations. (d) Corresponding variation of $\Delta T_c$  and $T_c$ with annealing time. (e) Thickness-dependent normalized resistance for films annealed at 830$^\circ$C and 860$^\circ$C. (f) Superconducting transition temperature $T_c$ as a function of hole concentration ($p$), exhibiting the characteristic dome-shaped dependence. Note: Except for subfigure (e), all subfigures present data from samples with a thickness of $\approx 110~\mathrm{nm}$.}
\label{RT}
\end{figure*}

We now turn towards a rigorous transport measurement analysis, where we notice that our samples span insulating phase to superconducting phase in a systematic manner by tuning the annealing conditions. While this has typically been achieved via electrostatic gating on very thin flakes \cite{Sterpetti2017,Sterpetti2019}, our results demonstrate that varying growth conditions has resulted in varying the doping level, thereby changing the resistance transition spectrum [Fig. 4 (a-b)].  The normalized resistance curves in Fig.~4(a) show that films annealed near 860$^\circ$C exhibit the highest $T_c$  and lowest normal-state resistance, indicating optimal carrier density and minimal disorder. In contrast, films annealed at lower temperatures display semiconducting behavior, while higher annealing temperatures suppress superconductivity due to oxygen loss and structural degradation. This behavior is quantitatively captured in Fig.~4(b), where both the  superconducting transition temperature ($T_c$) decreases and transition width ($\Delta~T_c$)  increases beyond the optimal annealing temperature, reflecting a shift toward the underdoped regime driven by reduced oxygen content. A similar non-monotonic dependence is observed in Fig.~4(c), where intermediate annealing durations yield highest $T_c$, whereas shorter durations result in insufficient crystallization and longer durations introduce disorder and compositional instability. The corresponding evolution of $\Delta~T_c$ and $T_c$ with annealing time in Fig.~4(d) further corroborates the above inference. The thickness dependence shown in Fig.~4(e) reveals that the obtained optimal conditions work the best for only 110 nm thickness; however, for different thicknesses optimal window can be different [refer to SI]. Finally, Fig.~4(f) demonstrates that the extracted $T_c$ follows the universal dome-like dependence [see SI for calculations] on hole concentration, consistent with established cuprate phase diagrams~\cite{Presland1991,Tallon1995,Sterpetti2017,Sterpetti2019,Wang2021,Raffy2007,Leng2012}. Together, these results establish that annealing-controlled oxygen stoichiometry governs carrier density and disorder, which in turn dictate superconducting phase coherence and transport properties \cite{Uemura1989,Homes2004}. A summary of the samples studied and their properties are shown in table I.

\begin{table}[h!]
\centering
\renewcommand{\arraystretch}{1.5}
\resizebox{\columnwidth}{!}{
\begin{tabular}{|c|c|c|c|c|c|c|}
\hline
Sample ID & Annealing Temp ($^\circ$C) & Annealing Time (h) & $T_c$ (K) & $\Delta T_c$ (K) & FWHM$_{2212}$ ($^\circ$) & $p$ (holes/Cu) \\
\hline
B59g & 830 & 3 & 0 & -- & - & -- \\
\hline
  B44a & 840 & 3 & 0 & -- & 0.2484 & --\\
\hline
  B44b & 850 & 3 & 0 & -- & 0.3036&--\\
\hline
  B44c & 860 & 3 & 69 & 6.554& 0.2398&0.16 \\
\hline
 B53d  &  870& 3 & 63 & 7.174 & 0.3186 & 0.128\\
\hline
 B47j & 880 & 3 & 49 & 9.588 & 0.321 & 0.101\\
\hline
  B47i & 860 & 0.5 & 0  & -- & 0.2448 & --\\
\hline
  B47h& 860 & 1  & 48 & 9.649 & 0.2841& 0.1\\
\hline
 B53f &860  & 2 & 53 & 11.163  & - & 0.107 \\
\hline
 B53e & 860 & 4 & 66 & 6.148 & - & 0.137 \\
\hline
 B72k & 860 & 5 & 58 & 7.636 & 0.2743 & 0.118\\
\hline
\end{tabular}
}
\caption{Sample parameters and superconducting properties}
\end{table}

\begin{figure}[t]
\centering
\includegraphics[width=14cm]{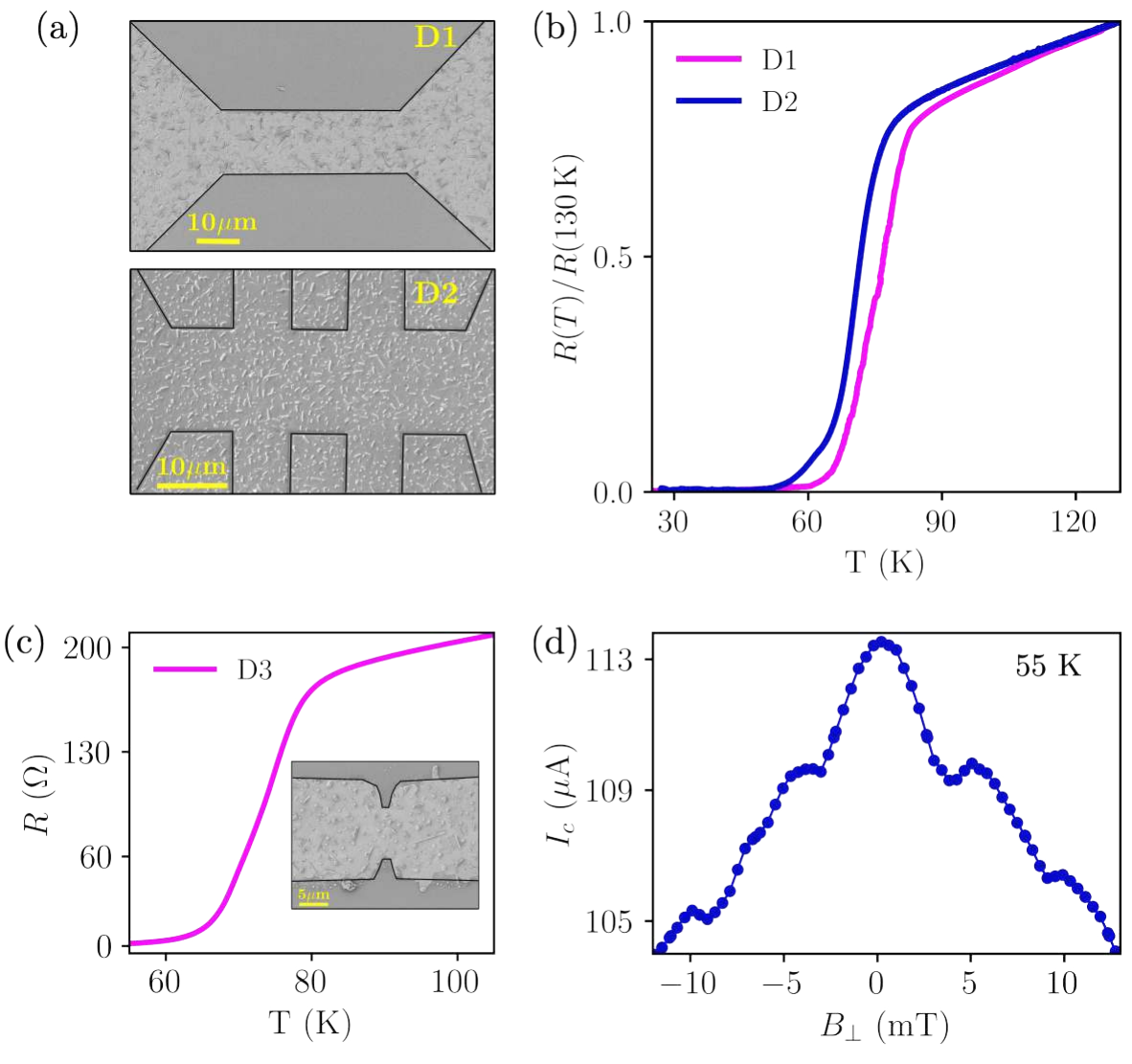}
\caption{(a) Scanning electron microscopy (SEM) images of BSCCO microbridge device patterned by optical lithography followed by a liftoff process (device D1, top panel) and a hallbar device patterned by focused ion beam (FIB) milling (device D2, bottom panel). Both microbridges have a width of $\approx 15\,\mu\mathrm{m}$ and a thickness of $\approx 110\,\mathrm{nm}$ and were fabricated under identical deposition and annealing conditions. (b) Corresponding normalized resistance as a function of temperature for devices D1 and D2. (c) $R(T)$ characteristics of the microbridge Josephson junction device D3, with a junction length of $\approx 1.5\,\mu\mathrm{m}$ and thickness $\approx 110\,\mathrm{nm}$. The inset shows the corresponding SEM image of the device. (d) Magnetic-field dependence of the critical current, $I_{\mathrm{c}}(B)$, measured for device D3 at $T=55\,\mathrm{K}$. }
\label{device}
\end{figure}

The dependence of $T_c$ on annealing temperature cannot be reduced to a
simple change in oxygen content. In BSCCO, oxygen incorporated into or
removed from the charge-reservoir layers changes their electrostatic
potential and the average Cu valence, and thereby controls charge transfer
to the $\mathrm{CuO_2}$ planes and increases the in-plane hole density $p$. Because $T_c$
depends non-monotonically on $p$, the effect of added oxygen depends on the
initial doping state.  Annealing also governs processes beyond oxygen doping. At low annealing
temperatures or short durations, limited cation and oxygen diffusion
suppresses the formation of the BSCCO-2212 phase and leaves poorly
connected superconducting regions. Within an intermediate annealing window,
enhanced atomic mobility promotes BSCCO-2212 phase formation, oxygen
redistribution, and improved connectivity between superconducting grains.
At higher temperatures or longer durations, oxygen loss, Bi volatility,
secondary-phase formation, interdiffusion, and surface coarsening increase
chemical and structural disorder. These effects produce spatial variations
in the local carrier density and pairing scale, which broaden the resistive
transition.

We therefore interpret annealing as jointly controlling phase formation,
oxygen redistribution, carrier doping, disorder, and inter-grain phase
coherence. The maximum $T_c$ observed near $860\,^{\circ}\mathrm{C}$ does
not simply mark the maximum oxygen content. It identifies a processing
window in which these competing effects are most favorably balanced.

Finally, the films studied here are not single-phase BSCCO-2212. X-ray diffraction
shows BSCCO-2212 together with a Bi-rich secondary phase that we assign to
$\mathrm{Bi_2O_3}$ [Fig.1 (a), Fig. 2 (a)].
Microscopy shows that this phase forms rod-like features distinct from the
surrounding BSCCO-2212 matrix [Fig.1 (d-g), Fig. 2 (d-g),].  Because the secondary phase is present across the whole annealing series,
the measured transport reflects the effective response of a spatially
inhomogeneous network rather than that of a chemically homogeneous
BSCCO-2212 phase. The secondary phase can locally alter oxygen chemical
potential, strain, and defect density, and it can interrupt conduction
paths. Conventional XRD cannot resolve the nanoscale distribution or local
stoichiometry of the phases. This is a limitation of the present work, and
phase-pure growth together with spatially resolved compositional analysis
is needed to establish the intrinsic doping dependence. Since the $\mathrm{Bi_2O_3}$ rods are spatially localized, the regions
containing them can in principle be avoided in device fabrication by
patterning devices on rod-free areas of the film.

Figure~5 extends the growth--property correlations established in Fig.~4 to the device level and demonstrates that the optimized BSCCO thin films are compatible with both bottom-up and top-down microfabrication routes. Broadly, superconducting micro- and nanodevices can be realized either by defining the device geometry prior to deposition (bottom-up) or by patterning an already grown film (top-down). Here, we demonstrate both approaches and establish fabrication protocols suitable for our sputtered BSCCO films. Figure~\ref{device} (a) shows microbridges with a nominal width of $15~\mu$m fabricated using the two methods. In the bottom-up approach (top panel), a photoresist is first spin-coated and lithographically patterned into the desired microbridge geometry. Following development, the BSCCO film is deposited by sputtering, lifted off, and subsequently annealed under the optimized conditions identified from the systematic growth study. In the top-down approach (bottom panel), an optimized and annealed continuous BSCCO film is directly patterned into a Hall-bar geometry using focused ion beam (FIB) milling. The resistance--temperature characteristics [figure~\ref{device} (b)] of the resulting devices confirm that superconductivity is retained following both fabrication routes. The transition temperatures differ by approximately $8$~K, with the FIB-patterned device exhibiting the lower $T_c$. Such suppression can plausibly arise from a combination of edge disorder, local oxygen redistribution, and Ga$^+$-ion implantation/damage produced during FIB milling. However, measurements over a statistically significant number of devices would be required to quantitatively establish the relative influence of the two fabrication processes. Accordingly, the present comparison is intended primarily to demonstrate the compatibility of sputtered BSCCO thin films with complementary microfabrication strategies rather than to establish the superiority of either method.

We further demonstrate a constriction-based, Dayem-bridge-like device fabricated using the lift-off process [Fig.~\ref{device} (c)]. The current--voltage characteristics were measured as a function of perpendicular magnetic field, $B_{\perp}$ (see Supplementary Information), and the switching current $I_c$, defined as the current at which the device switches from the superconducting to the resistive state, was extracted at each field. The resulting $I_c(B_{\perp})$ [Fig.~\ref{device} (d)] exhibits reproducible oscillations with a characteristic field period of approximately $\Delta B \simeq 4$~mT, consistent with magnetic-flux interference through an effective weak-link region. For a Josephson weak link, successive minima of an ideal Fraunhofer interference pattern occur upon the addition of approximately one superconducting flux quantum,
\begin{equation}
\Phi_0 = \frac{h}{2e} = 2.068\times10^{-15}~\mathrm{Wb},
\end{equation}
such that
\begin{equation}
\Delta B A_{\mathrm{eff}} \simeq \Phi_0.
\end{equation}
The experimentally observed periodicity therefore gives
\begin{equation}
A_{\mathrm{eff}}
\simeq \frac{\Phi_0}{\Delta B}
= \frac{2.068\times10^{-15}~\mathrm{Wb}}
{4\times10^{-3}~\mathrm{T}}
\simeq 5.2\times10^{-13}~\mathrm{m}^2
\simeq 0.52~\mu\mathrm{m}^2.
\end{equation}
For comparison, a geometrical junction area of $1~\mu\mathrm{m}^2$ would correspond to a flux-quantum field scale of
\begin{equation}
\Delta B = \frac{\Phi_0}{A}
\simeq 2.07~\mathrm{mT}.
\end{equation}
Thus, the measured $\sim4$~mT periodicity indicates that the electrically active weak-link area is smaller than the optically defined constriction area. Such a reduction can originate from spatially nonuniform superconductivity within the constriction, current crowding, edge disorder, or compositional/structural inhomogeneity, including Bi-rich regions, which can restrict the effective supercurrent-carrying cross-section. The effective magnetic area can additionally differ from the lithographically defined area because of flux focusing and finite magnetic-field penetration into the superconducting electrodes.

The observation of a field-modulated switching current at temperatures as high as $55$~K provides evidence for phase-coherent weak-link transport in the patterned sputtered BSCCO film and represents an important step beyond merely retaining superconductivity following device fabrication. While an unambiguous identification of an ideal Josephson Fraunhofer response would require detailed analysis of the complete lobe structure, its symmetry and reproducibility, the experimentally observed characteristic field scale is quantitatively consistent with the modulation of the weak-link supercurrent by approximately one superconducting flux quantum per oscillation. These results therefore establish a pathway from scalable thin-film growth to lithographically defined high-$T_c$ superconducting weak links and provide a basis for developing more complex superconducting circuits based on sputtered BSCCO thin films.

\begin{figure*}[t]
\centering
\includegraphics[width=\textwidth]{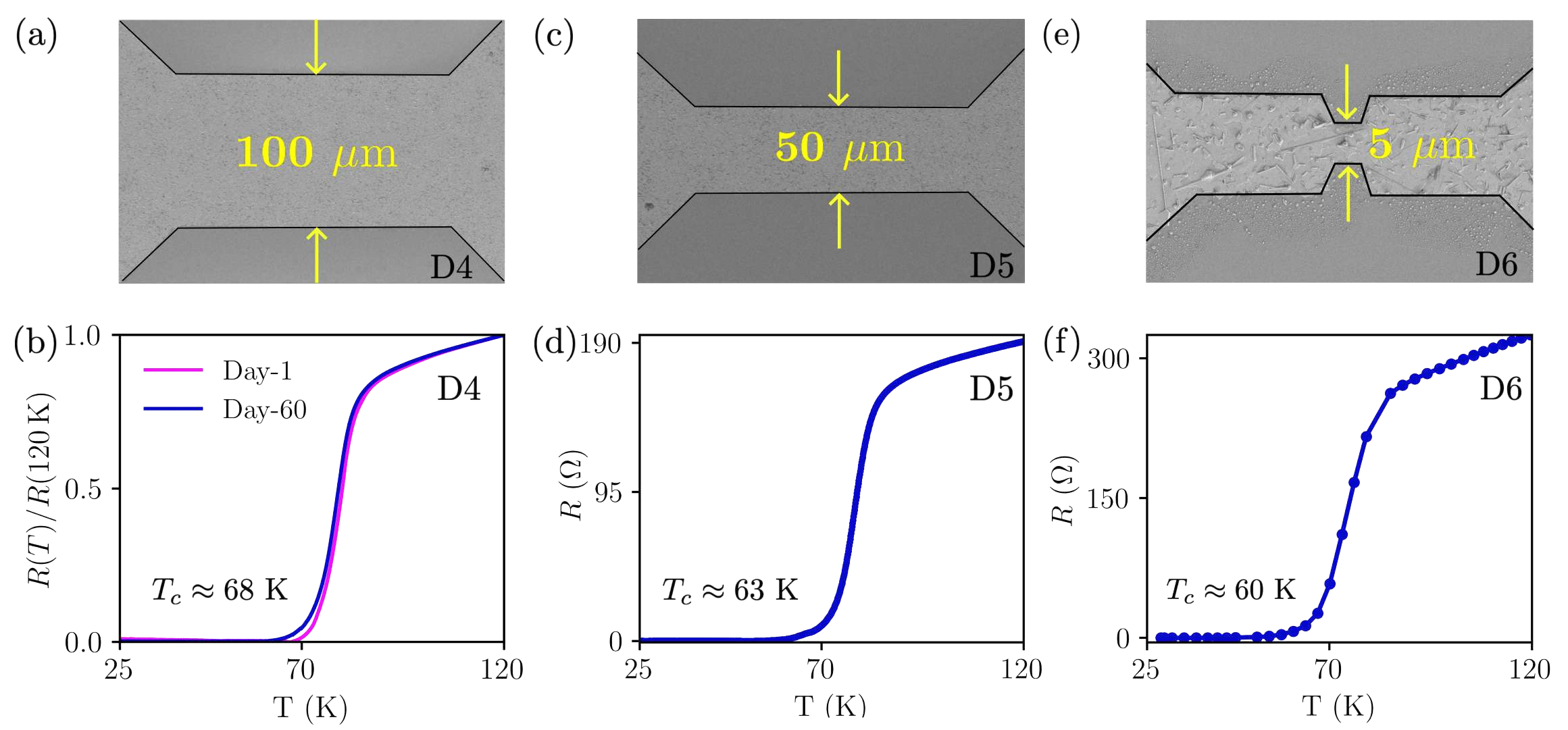}
\caption{(a, c, e) Scanning electron microscopy (SEM) images of microbridges of BSCCO thin film with widths of $\approx 100\,\mu\mathrm{m}$, $\approx 50\,\mu\mathrm{m}$, and $\approx 5\,\mu\mathrm{m}$, respectively. All microbrdige devices have a thickness of $\approx 110\,\mathrm{nm}$ and were fabricated under identical deposition and annealing conditions. (b, d, f) Corresponding $R$--$T$ characteristics of the respective microbridges.}
\label{device2}
\end{figure*}

To study the variation of $T_c$ with inhomogeneity, we fabricated micro-bridges of varying lateral widths of $\approx 100\,\mu\mathrm{m}$, $\approx 50\,\mu\mathrm{m}$, and $\approx 5\,\mu\mathrm{m}$ and measured their $T_c$ [Fig.~\ref{device2} (a-f)]. All the devices have a thickness of $\approx 110\,\mathrm{nm}$ and were fabricated under identical deposition and annealing conditions. The $T_c$ remains largely stable across device sizes, though larger devices exhibit marginally higher $T_c$, as expected. Narrower devices show relatively lower $T_c$ and broader transition widths, which may be attributed to edge disorder. Furthermore, our devices demonstrate excellent air stability: the same device [D4 - Fig.~\ref{device2} (a, b)], measured 60 days apart, shows nearly identical $T_c$. This represents a significant advantage over exfoliated devices, which suffer from poor air stability owing to their much larger surface-to-volume ratio exposed to air compared to our films.

The combined structural, chemical, and transport analyses presented in Figs.~1–6 establish a comprehensive and self-consistent optimization of Bi$_2$Sr$_2$CaCu$_2$O$_{8+\delta}$ thin film growth, wherein thermodynamic and kinetic parameters are systematically tuned to achieve precise control over phase formation, oxygen stoichiometry, and superconducting functionality [Table I]. These results establish a unified framework in which growth temperature and annealing duration control the defect landscape and oxygen content. This in turn dictates carrier concentration, superconducting phase coherence, and device performance. The demonstrated ability to reproducibly engineer  superconducting BSCCO thin films, and to translate these properties into operational devices, represents a significant step toward the realization of oxide-based quantum devices. These findings provide a practical framework for integrating cuprate superconductors into scalable electronic and quantum devices.

DS thanks IISc start-up grant, Ministry of Electronics and Technology, Indian Space Research Organization for funding. Authors duly acknowledge funding from INOXCVA and INOX Airproducts for funding via CSR grants. PB thanks Anusandhan National Research Foundation, National Postdoctoral fellowship (PDF/2023/000444) for financial support. Authors are grateful to micro and nano characterization facility, CeNSE and national nanofabrication facility, CeNSE for facilities usage.

\bibliography{references}

\clearpage
\onecolumn
\includepdf[pages=-,pagecommand={\thispagestyle{empty}}]{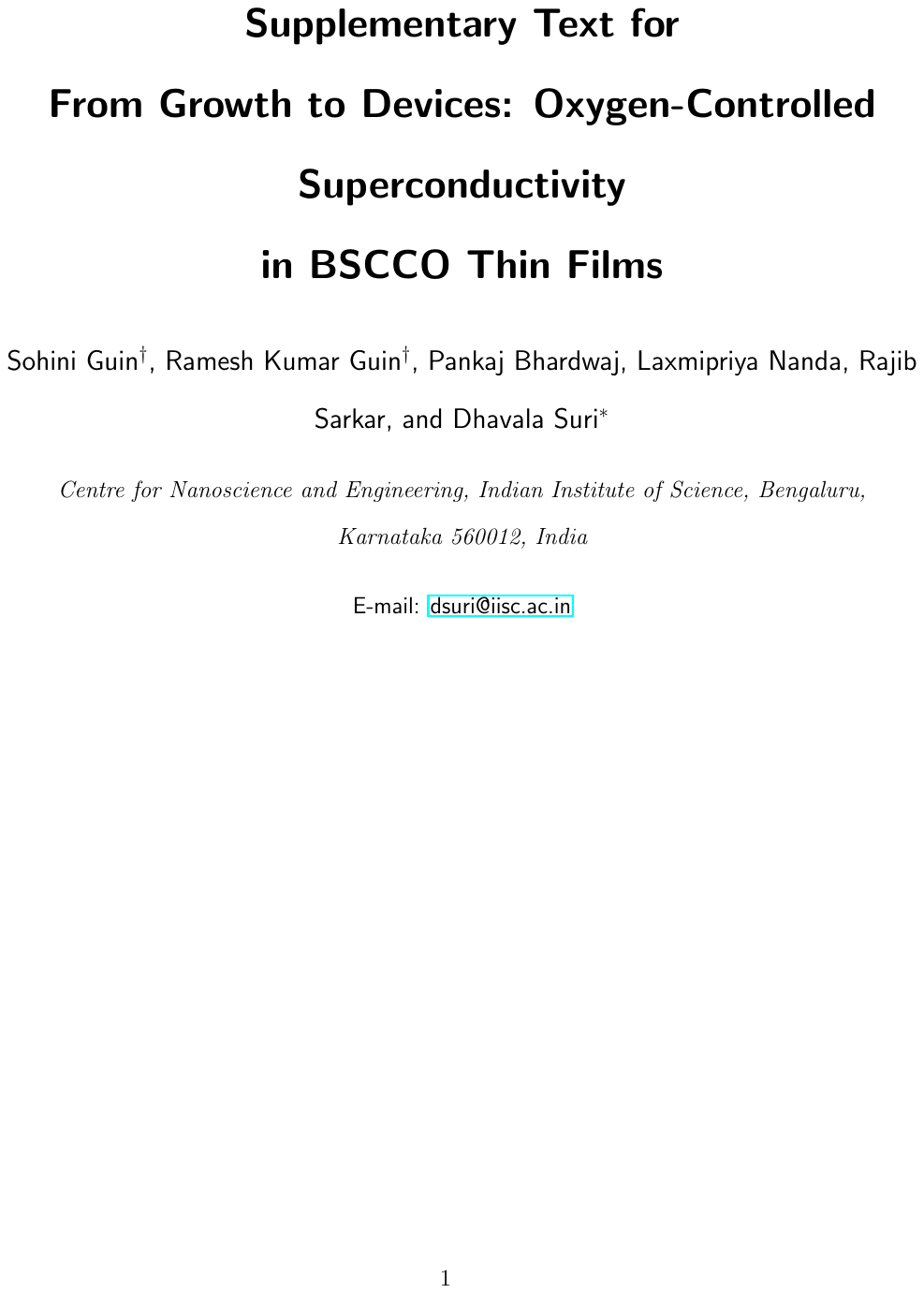}

\end{document}